\documentclass[twocolumn,amsmath,amssymb]{revtex4-1}

\usepackage{graphicx}% Include figure files
\usepackage{dcolumn}% Align table columns on decimal point
\usepackage{bm}% bold math

\begin{document}
\title{Hidden Variable Theory of a Single World from Many-Worlds Quantum Mechanics}
\author{Don Weingarten}
\affiliation{donweingarten@hotmail.com}

%\receipt{}
\begin{abstract}
We propose a method for finding an initial state vector which by ordinary Hamiltonian time evolution follows a single branch of many-worlds quantum mechanics. The resulting deterministic system
appears to exhibit random behavior as a result of the successive emergence over time of information
present in the initial state but not previously observed.
\end{abstract}
%\pacs{03.65.Ca, 02.30.Cj, 03.65.Ta}

\maketitle
\section{\label{sec:intro}Introduction}

Microscopic particles have wave functions spread over all possible positions. 
Macroscopic objects simply have positions, or at least center-of-mass positions.
How to apply the mathematics of quantum mechanics to extract predictions 
registered in the macroscopic world of positions 
from experiments on microscopic systems having wave functions but not definite positions 
is well understood for all practical purposes.
But less well understood, or at least not a subject on which there is a clear consensus, is 
how in principle the definite positions of the macroscopic world emerge from the microscopic 
matter of which it is composed, which has only wave funtions but not definite positions.
There is a long list of proposals.
In the present article we will add yet one more entry to the list.  

We begin in Section \ref{sec:problems} with a brief reminder of the ``problem of measurement'' which arises
for an experiment
in which a microscopic system interacts with an external macroscopic measuring device 
with both systems assumed governed by quantum mechanics.
We then review
the many-worlds interpretation's \cite{Everett, DeWitt} proposed
resolution of this problem.
The many-worlds interpretation, however, has well-known difficulties of its own.
We next summarize how these problems are, to some degree, resolved by taking into account
the process of environmentally-induced decoherence \cite{Zeh, Zurek1, Zurek2, Schlosshauer, Wallace}.
But problems still remain. 
In Section \ref{sec:dqm} we introduce a
hypothesis on the mathematical structure of environmentally-induced decoherence. 
Subject to this hypothesis, we propose a formulation of quantum mechanics
which we believe resolves the difficulties in the initial combination of
many-worlds and decoherence.
The proposal incorporates purely
deterministic 
time development by the usual unitary operator following, however,
only a single branch of many-worlds quantum mechanics.
The system's initial state determines which branch is followed and is chosen
from a particular random ensemble.
In Section \ref{sec:born} we show that the probability
the random ensemble assigns each branch agrees with
the probability that would be assigned to each branch by
the Born rule. As a consequence the proposed formulation
of quantum mechancs predicts the same observable 
results for experiments as would be obtained from 
the Born rule.
In Section \ref{sec:toy} 
we present a toy model which illustrates how the proposed
deterministic theory works.
In Section \ref{sec:bell} we 
apply the deterministic theory to 
a Bell's theorem test and obtain,
as required by the results of Section \ref{sec:born}, 
the predictions expected of quantum mechanics
and in violation of the constraints implied by Bell's theorem for a certain class of hidden variable theories.
We conclude in Section \ref{sec:states} with several comments about the random ensemble
from which initial states are drawn.

\section{\label{sec:problems}Problems}
Let $\mathcal{S}$ be a microscopic system to be measured, with corresponding state space $\mathcal{H}_\mathcal{S}$
for which a basis is $\{ |s_i>\}, i > 0$. Let 
$\mathcal{M}$ be a macroscopic measuring device with corresponding state space $\mathcal{H}_\mathcal{M}$
containing the set of vectors $\{ |m_i>\}, i \geq 0$.  For each different value of $i > 0$ the state $|m_i>$ is 
a macroscopically distinct meter reading.
Let $|m_0>$ be an initial state showing no reading. In the 
combined system-meter product state space 
$\mathcal{H}_\mathcal{S} \otimes \mathcal{H}_\mathcal{M}$ a measurement of $\mathcal{S}$ by $\mathcal{M}$ over some time interval
takes each possible initial state $|s_i> |m_0>$ into the corresponding final state $|s_i>|m_i>$
with the measuring device displaying the measured value of the microscopic system's variable. 
By linearity of quantum mechanical time evolution, however, it then follows that a measurement
with a linear superpostion in the initial state will yield a final state also with a superposition
\begin{multline}
(\frac{1}{\sqrt{2}}|s_1> + \frac{1}{\sqrt{2}} |s_2>) |m_0> \rightarrow \\ 
\frac{1}{\sqrt{2}}|s_1> |m_1> + \frac{1}{\sqrt{2}} |s_2> |m_2>. 
\end{multline}
In the measured final state, the meter no longer has a single value, but a
combination of two values which cannot, by itself,  be connected to a recognizable configuration of a macroscopic object.
The absence of a recognizable configuration for the macroscopic device is the ``problem of measurement''.

The resolution of this problem proposed by the many-worlds interpretation of quantum mechanics \cite{ Everett, DeWitt} 
is that the states $|s_1> |m_1>$ and $|s_2> |m_2>$ actually represent two different worlds. In each world the 
meter has a definite position but with different positions in the two different worlds. Among the problems of this proposed
resolution is that 
\begin{multline}\label{basis}
\frac{1}{\sqrt{2}} |s_1> |m_1> + \frac{1}{\sqrt{2}} |s_2> |m_2> = \\
\frac{1}{\sqrt{8}} (|s_1> + |s_2>)(|m_1> + |m_2>) + \\
\frac{1}{\sqrt{8}} (|s_1> - |s_2>)(|m_1> - |m_2>). 
\end{multline}
Thus again two worlds, but now with meter configuration of $|m_1> + |m_2>$ in one and
$|m_1> - |m_2>$ in the other and thus a recognizable configuration for a macroscopic object in neither.
According to which basis of $\mathcal{H}_\mathcal{M}$ should the two worlds be split? 
Without some reason for a choice, many-worlds quantum mechanics still has a problem, the ``preferred basis problem''.

A resolution to this problem is proposed to occur through environmentally-induced decoherence 
\cite{Zeh, Zurek1, Zurek2, Schlosshauer, Wallace}.
According to this proposal, in abbreviated summary, the system-meter combination should not be considered in isolation
but instead an account is required of the rest of the macroscopic environment with which the meter can interact.
When the value of a macroscopic meter is changed by recording the value of a microscopic coordinate, the meter's new state
rapidly becomes entangled with a large number of other degrees of freedom in the environment 
\begin{multline}
\label{entangled}
\frac{1}{\sqrt{2}}|s_1> |m_1> + \frac{1}{\sqrt{2}} |s_2> |m_2> \rightarrow \\
\frac{1}{\sqrt{2}}|s_1'> |m_1'> |e_1> + \frac{1}{\sqrt{2}} |s_2'> |m_2'> |e_2>.
\end{multline}
For a particlar choice of basis, determined by the system's dynamics, entanglement
of the meter with the environment
is optimal in the sense that $|s_1'> |m_1'>$ and $|s_2'> |m_2'>$ are 
nearly equal to $|s_1> |m_1>$ and $|s_2> |m_2>$, respectively, 
$| e_1>$ and $| e_2>$ are nearly orthogonal, the process of the
meter becoming entangled with the environment proceeds as quickly as possible,
and $| e_1>$ and $| e_2>$ almost do not mix in the course of futher time development
becoming, as a result, permanent records of the measurement results.
In addition, $|e_1>$ and $|e_2>$ include many redundant copies
of the information in $|s_1> |m_1>$ and $|s_2> |m_2>$.
Based on these various considerations it is argued that the coordinates with respect
to the optimal basis behave essentially as classical degrees of freedom.
Correspondingly, for many-worlds augmented with decoherence, the cirumstance 
under which a system splits into distinct worlds is when a superposition
has been produced mixing distinct values of one of these effectively classical 
degrees of freedom. Each distinct value of the coordinate in such a superposition
goes off into a distinct world. 

But a variety of problems remain for the combination of decoherence and many-worlds.
In particular, 
environmentally-induced decoherence occurs over some finite time interval, no matter how short,
occurs over some extended region of space,   
the entangled states $|s_1'> |m_1'>$ and  $|s_2'> |m_2'>$
are not exactly equal to 
$|s_1> |m_1>$ and $|s_2> |m_2>$, respectively, and $|e_1>$ and $|e_2>$ are not exactly orthogonal.
When, exactly, over the time interval of decoherence, does the splitting of the world in two parts 
occurr? And since the process extends over space, this timing will differ in different
Lorentz frames. Which is the correct choice?  In addition, how accurate do the 
approximate equalities and approximate orthogonality in the entanglement process
have to be to drive a split? 
These questions may be
of no practical consequence in treating the meter readings as classical
degrees of freedom after entanglement and using the resulting values to formulate observable predictions.
For any possible choice of the location of these boundaries, however, 
there will be at least some events in the vastness of space and time
that either do or do not lead to new branches of the universe depending on essentially arbitrary choices of
the parameters distinguishing acceptable from unacceptable entanglement events.
These problems seem hard to square with the view that the
state vector is the fundamental substance of reality and
hard to accept as features of a process in which this
fundamental substance splits into distinct copies.
While it might be possible to argue that these 
elements of fuzziness in splitting events are necessary 
consequences of the quantum nature of the world, 
it still seems they would be better avoided if possible.

\section{\label{sec:dqm}Deterministic Quantum Mechanics}

We now consider a modified version of many-worlds plus decoherence which avoids these
three issues by pushing the process of world splitting infinitely far back into the
past, or at least back before any potential world splitting events occur. 
In place of splitting, each possible world will be specified by a different
initial state drawn from a random ensemble.

Choose some reasonable but otherwise arbitrary way of resolving the event time ambiguities.
Let $s_i$ be the set of possible outcomes of splitting event $i$, occuring
at time $t_i$. We assume, for simplicity, only a finite sequence of $N$ such
splittings, the last occuring at time $t_N$.
Let $h$ specify a sequence of splitting results $( r_0, r_1, ... r_k)$, up to some time $t_k$, 
with each $r_i \in s_i$. 
Let $\mathcal{S}_k$ be the set of all such possible sequences for events up to and including
time $t_k$.
Let $\mathcal{H}$ be the space of states of the universe.
Let $| \Psi( t) > \in \mathcal{H}$ be the 
state of the universe at time $t$
\begin{equation}\label{evolution}
|\Psi( t)> = exp( -i H t) | \Psi(0)>
\end{equation}
for system Hamiltonian $H$.
                   
We adopt from studies of decoherence \cite{Zeh, Zurek1, Zurek2, Schlosshauer}
the hypothesis that each macroscopic splitting event,
after entanglement with the environment, 
leaves a permanent footprint in the degrees of freedom of $\mathcal{H}$,
consisting of multiple copies of the information which was recorded
in the event driving the split.
Since each record is assumed permanent, records accumlate over time
so that
at any $t$ shortly after event time $t_k$ and before $t_{k+1}$, the
state vector can be expressed as a sum over branches 
of splitting events of the form
\begin{equation}\label{entangled1}
|\Psi( t)> = \sum_{h \in \mathcal{S}_k} | \Psi( h, t)>,
\end{equation}
where $|\Psi( h, t)>$ a vector for which the environmental degrees 
of freedom carry a record of
splitting history $h$,
the sum is over splitting histories up to time $t_k$,
and environmental record components of $|\Psi( h, t)>$ and $|\Psi( h', t)>$ for distinct $h$ and $h'$
are orthogonal, or nearly so.
The decomposition in Eq. (\ref{entangled1}) will hold
up until shortly after the next splitting time $t_{k + 1}$, at 
which time an updated expression with histories $h$
drawn from $\mathcal{S}_{k+1}$ will take over.

From Eq. (\ref{entangled1}) applied 
at some time $t$ after the last splitting
event $t_N$, for some history $h \in \mathcal{S}_N$, define
from $|\Psi( h, t)>$
its evolution back to time 0
\begin{equation}
\label{onebranch2}
|\Psi( h, 0)>  =  exp( i H t) | \Psi( h, t)>.
\end{equation}
If $| \Psi( h, 0)>$ is now evolved forward again in time retracing its history up to $t$,
the result at each $t_k$ has to follow the branch $r_k$ specified in $h$.
If $| \Psi( h, 0)>$ is evolved forward again in time up to $t$
the result will clearly have to return to $|\Psi( h, t)>$ exhibiting splitting history $h$.
But following any splitting event at some $t_k$, the contribution
$r_k$ of that event to the history $h$ is permanently recorded.
Thus the only way $|\Psi( h, t)>$ can aquire a record of the full history  
$h$ as of time $t$ is if each $r_k$ is correctly recorded at $t_k$ and therefore the
split at $t_k$ occurs as specified by $r_k$. 

The system's initial state we now assume is found
by a single random choice at time 0
from an ensemble of $| \Psi( h, 0)>$ for $h$ drawn from $\mathcal{S}_N$ with each having
probability $< \Psi(h, 0) | \Psi(h, 0)>$.
The result is that each history $h$ will occur with
the same probability that would be found
if splitting events were treated as sequences of observations
by some external observer and their 
probabilities then calculated by the Born rule.
A proof of this statement will be given in Section \ref{sec:born}. 

Each $| \Psi(h, t)>$ may be viewed as a different world.
But splitting between the different worlds occurs only once, at
the beginning of history. 

One piece of the formulation of quantum mechanics
we now propose remains approximate, but another has become exact.
What has become exact is the time evolution by Hamiltonian $H$ 
of the state vector $| \Psi( h, 0)>$ drawn from the initial
random ensemble. This state vector we take as the underlying
substance of the real world. What is approximate, on the other hand,
is the extent to which this evolution follows splitting 
history $h$. This tracking is approximate both because
the time at which the entanglement equation Eq. (\ref{entangled1}) holds 
is approximate
and because the orthogonality of the environmental record coponents of
vectors $| \Psi( h, t)>$ and $| \Psi( h', t)>$ for distinct $h$ and $h'$ is
approximate. Put differently, the underlying
substance of the microscopic world follows an exact law, 
to which the macroscopic description is an approximation.
By contrast, the combination of many-worlds and decoherence
considered in Section \ref{sec:problems} included
only an approximate macroscopic world as the substance of reality.

The formulation we propose is a kind of hidden variable theory,
with the hidden variables present in the initial $| \Psi( h, 0)>$.
They emerge in macroscopic reality only over time through their
influence on the sequence of splitting results $( r_0, r_1, ... r_N)$ 
forming the history $h$.

\section{\label{sec:born}Born Rule}

We will show that the probability weight $< \Psi(h, 0) | \Psi(h, 0)>$ assigned
to $| \Psi( h, 0)>$ in the initial random ensemble is the same as the
probability which the Born rule would assign to a series
of observations of the system by an outside observer at splitting
event times $(t_0, t_1, ...t_N)$ yielding the sequence of results
$(r_0, r_1, ...r_N)$ of the history $h$.

By the unitarity of time evolution, we can restore $< \Psi( h, 0)| \Psi( h, 0)>$ to a time $t$
shortly after $t_N$
\begin{equation}
\label{backtofuture}
< \Psi(h, 0) | \Psi(h, 0)> = < \Psi(h, t) | \Psi(h, t)>.
\end{equation}
But since $h$ exhibits result $r_N$ at $t_N$
we have
\begin{subequations}
\label{firststep}
\begin{eqnarray}
| \Psi(h, t)> & = & P( r_N) | \Psi(h, t)> \\
& = & P( r_N) exp[ -i H (t_N - t_{N - 1}] \times \\
& &| \Psi( h, t_{N-1})> 
\end{eqnarray}
\end{subequations}
where $P( r_N)$ is the projection operator onto the subspace of $\mathcal{H}$ 
for result $r_N$.

Iterating the argument for Eqs. (\ref{firststep}), we obtain
\begin{subequations}
\label{nextstep}
\begin{eqnarray}
| \Psi(h, t)> & = & P( r_N) exp[ -i H (t_N - t_{N - 1})]  \times \\
& & ... P( r_1) exp[ -i H (t_1 - t_0]  \times \\
& &     P( r_0) exp( -i H t_0)  | \Psi( h, 0)> \\
& \equiv & Q( h) | \Psi( h, 0)>, 
\end{eqnarray}
\end{subequations}
where for notational convenience we define $Q( h)$ to be
the preceding product of projections and time evolution.
For any $h'$ which differs from $h$ by at least one result
$r_i$, however, we have
\begin{equation}
\label{finalborn}
Q(h) |  \Psi(h', 0)> = 0,
\end{equation}
since if Eq. (\ref{nextstep}) is unfolded for $| \Psi(h', 0)>$, at step $r'_i$ 
the projection $P( r_i)$ will annihilate the evolving image of $| \Psi(h', 0)>$. 

From Eqs. (\ref{entangled1}, \ref{onebranch2}), we have
\begin{equation}\label{entangledagain}
|\Psi( 0)> = \sum_{h \in \mathcal{S}_N} | \Psi( h, 0)>.
\end{equation}
Combining Eqs. (\ref{backtofuture} - \ref{entangledagain}) gives
\begin{equation}
\label{finalfinal}
< \Psi(h, 0) | \Psi(h, 0)> = < \Psi(0) | Q(h)^\dagger Q(h)| \Psi(0)>.
\end{equation}
The expression on the right hand side of Eq. (\ref{finalfinal}), as
advertised, is the
probability which the Born rule would assign to a series
of observations of the system by an outside observer at splitting
event times $(t_0, t_1, ...t_N)$ yielding the sequence of results
$(r_0, r_1, ...r_N)$ of the history $h$.

A consequence of Eq. (\ref{finalfinal}) is that the proposed formulation
of quantum mechancs predicts the same observable 
results for experiments as would be obtained from 
the Born rule.

\section{\label{sec:toy}Toy Model}

We now turn to a toy model.
Consider a universe composed of a spin $s$ taking two values
$| \uparrow>$ and $| \downarrow>$ of z-direction spin which interacts
with an environment composed of variables
$e_0$, $e_1$, $e_2$, and $e_3$ each
taking values, $|0>$ and $| 1>$. Except for transitions to be
specified below at times $t_1$ and $t_2$, we assume a Hamiltonian
diagonal in the basis $| e_0> |e_1> |e_2> |e_3>$ so that
\begin{multline}
\label{hamiltonian}
exp( -i H t) | e_0> |e_1> |e_2> |e_3> = \\
 exp( -i t \sum \omega_i e_i) | e_0> |e_1> |e_2> |e_3> \equiv \\
 | e_0, t> |e_1, t> |e_2, t> |e_3, t>.
\end{multline}
The $\omega_i$ we assume chosen in such a way that none of the states $| e_0, t> |e_1, t> |e_2, t> |e_3, t>$
are degenerate. The result of these choices will be that the
$| e_0, t> |e_1, t> |e_2, t> |e_3, t>$
form an optimal environmentally selected memory basis for the events at $t_1$ and $t_2$.

Assume an initial state $|\Psi( 0)>$ at time 0
of $| \uparrow> |0, 0> |0, 0> | 0, 0> | 0, 0>$.
From time 0 to some later $t_1$ the system evolves according to Eq. (\ref{hamiltonian}).
Then at $t_1$ the z-direction spin of $s$ is recorded by
$e_0$ and $e_1$ according to the unitary time step $U_1$ 
\begin{subequations}
\label{U_1}
\begin{multline}
\label{U_1a}
U_1 |\uparrow>| e_0, t> | 0, t> | e_2, t> | e_3, t>  =   \\ |\uparrow> | e_0, t> | 1, t> | e_2, t> | e_3, t> 
\end{multline}
\begin{multline}
\label{U_1b}
U_1 |\uparrow>| e_0, t> | 1, t> | e_2, t> | e_3, t>  =  \\ |\uparrow> | e_0, t> | 0, t> | e_2, t> | e_3, t> 
\end{multline}
\begin{multline}
\label{U_1c}
U_1 |\downarrow>| 0, t> | e_1, t> | e_2, t> | e_3, t>  = \\ |\downarrow> | 1, t> |e_1 , t> | e_2, t> | e_3, t> 
\end{multline}
\begin{multline}
\label{U_1d}
U_1 |\downarrow>| 1, t> | e_1, t> | e_2, t> | e_3, t>  = \\ |\downarrow> | 0, t> |e_1 , t> | e_2, t> | e_3, t> 
\end{multline}
\end{subequations}
which hold for any choice of $e_0, ... e_3$.
Upon completion of this process, the system's state becomes $| \uparrow>| 0, t_1> | 1, t_1> |0, t_1> | 0, t_1>$
The sytem again follows Eq. (\ref{hamiltonian}) until some later time $t_2$ at which
the x-direction spin of $s$ is recorded on $e_2$ and $e_3$  
according to the unitary time step $U_2$ 
which is the same as $U_1$ in Eqs. (\ref{U_1}) but with $e_0$ and $e_1$ replaced, respectively, by
$e_2$ and $e_3$ and with the spin states $|\uparrow>$, $|\downarrow>$
of $s$ replaced, respectively, by $|\uparrow>_x$, $|\downarrow>_x$
\begin{subequations}
\begin{eqnarray}
|\uparrow>_x & = & \frac{1}{\sqrt{2}}|\uparrow> + \frac{1}{\sqrt{2}} |\downarrow>, \\ 
|\downarrow>_x & = & \frac{1}{\sqrt{2}}|\uparrow> - \frac{1}{\sqrt{2}} |\downarrow>.
\end{eqnarray}
\end{subequations}
Upon completing this step, the system's state becomes
\begin{multline}
\frac{1}{2} |\uparrow> | 0, t_2>  | 1, t_2> | 0, t_2> | 1, t_2> + \\ \frac{1}{2} |\downarrow> | 0, t_2>  | 1, t_2> | 0, t_2> | 1, t_2>  + \\
\frac{1}{2} |\uparrow> | 0, t_2> | 1, t_2> | 1, t_2> | 0, t_2> - \\ \frac{1}{2} |\downarrow>  | 0, t_2> | 1, t_2> | 1 , t_2> | 0, t_2>. 
\end{multline}

Histories can be labeled with the values of $( e_0, e_1, e_2, e_3)$.
The two non-zero results left after $t_2$ are $( 0, 1, 0, 1)$ and $( 0, 1, 1, 0)$.
The random ensemble of states
at time $t_2$ thus consists of the two vectors
\begin{subequations}
\begin{multline}  
\frac{1}{\sqrt{2}}| \uparrow> | 0, t_2> |1, t_2> |0, t_2> |1, t_2> + \\ \frac{1}{\sqrt{2}}|\downarrow>  | 0, t_2> |1, t_2> |0, t_2> |1, t_2>, 
\end{multline}
\begin{multline}
\frac{1}{\sqrt{2}}| \uparrow> | 0, t_2> |1, t_2> |1, t_2> |0, t_2> -\\ \frac{1}{\sqrt{2}}|\downarrow>  | 0, t_2> |1, t_2> |1, t_2> |0, t_2>,
\end{multline}
\end{subequations}
each with probability of $\frac{1}{2}$. 

The ensemble of initial states $|\Psi( h, 0)>$
can be recovered by applying
the reversed time evolution operators $U_2^\dagger$, $U_1^\dagger$ combined with the reverse of Eq. (\ref{hamiltonian}). 
The result is 
\begin{subequations}
\label{initial}
\begin{multline}  
\frac{1}{\sqrt{2}}| \uparrow>|0, 0> |0, 0> 0, 0> |0, 0> + \\ \frac{1}{\sqrt{2}}|\downarrow>|1, 0> |1, 0> 0, 0> |0, 0> , 
\end{multline}
\begin{multline}
\frac{1}{\sqrt{2}}| \uparrow>|0, 0> |0, 0> 0, 0> |0, 0> - \\ \frac{1}{\sqrt{2}}|\downarrow> |1, 0> |1, 0> 0, 0> |0, 0>. 
\end{multline}
\end{subequations}
A repeat of time evolution forward shows that the first of these states produces an
$(e_0, e_1, e_2, e_3)$ history which is purely $ (0, 1, 0, 1)$ and the second produces
a history which is purely $( 0, 1, 1, 0)$. 

\section{\label{sec:bell}Bell's Theorem}

For a system $| \Psi>$ consisting of two spin $\frac{1}{2}$ particles in a total angular
momentum 0 state, the standard formulation of quantum mechanics yields
\begin{equation}
\label{standard}
< \Psi | ({\vec{\sigma}^1} \cdot \hat{u}) ({\vec{\sigma}^2} \cdot \hat{v}) | \Psi> = - {\hat{u}} \cdot {\hat{v}},
\end{equation}
where $\vec{\sigma}^i$ is the vector of sigma matrices acting on spin $i$ and $\hat{u}$ and 
$\hat{v}$ are two unit vectors. Bell's theorem is that the 
result of Eq. (\ref{standard}) can not be realized in a particular
class of hidden variable theories.

We now consider an experimental setup for deterministic quantum mechanics which
measures the expectation value in Eq. (\ref{standard}). As expected
from the proof in Section \ref{sec:born}, the results predicted by the
determinstic formulation of quantum mechanics will reproduce
Eq. (\ref{standard}). For convenience, we will take $\hat{u}$ to be
in the z-direction and $\hat{v}$ to be rotated from $\hat{u}$ by
some angle $\theta$.

We assume a system composed of a large number of identical subsystems 
$\mathcal{W}_1, ... \mathcal{W}_N$. 
Each subsystem includes two spins
$s_0$ and $s_1$, each $s_i$ takes two values
$| \uparrow>$ and $| \downarrow>$ of z-direction spin.
Each subsystem includes also 
an environment composed of variables
$e_0$, $e_1$, $e_2$, and $e_3$ each
taking values, $|0>$ and $| 1>$. As in Section \ref{sec:toy},
except for transitions to be
specified at a sequence of times $t_i$, we assume a Hamiltonian
diagonal in the basis $| e_0> |e_1> |e_2> |e_3>$ 
according to Eq. (\ref{hamiltonian}), with corresponding vectors
$| e_0, t> |e_1, t> |e_2, t> |e_3, t>$
forming an optimal environmentally selected memory basis for the event at each $t_i$.

Assume an initial state $|\Psi( 0)>_i$ for subsystem $\mathcal{W}_i$ at time 0
of 
\begin{multline}
\label{subsystem}
| \Psi(0)>_i = \frac{1}{\sqrt{2}}(| \uparrow> |\downarrow> - | \downarrow> |\uparrow>) \times \\
|0, 0> |0, 0> | 0, 0> | 0, 0>.
\end{multline}

For each subsystem $\mathcal{W}_i$ from time 0 to some later $t_i$, time evolution follows Eq. (\ref{hamiltonian}).
Then at $t_i$ the z-direction spin of $s_0$ is recorded by
$e_0$ and $e_1$ according to the unitary time step $U_1$ of Eqs. (\ref{U_1}).
In addition, at the same time, 
the spin of $s_1$ is recorded on $e_2$ and $e_3$ along
a direction rotated by some angle $\theta$ from the z-direction
toward the x-direction,   
according to the unitary time step $U_2$. Time step $U_2$ 
is the same as $U_1$ in Eqs. (\ref{U_1}) but with $e_0$ and $e_1$ replaced, respectively, by
$e_2$ and $e_3$ and with the spin states $|\uparrow>$, $|\downarrow>$
of $s$ replaced, respectively, by $|\uparrow>_{\theta}$, $|\downarrow>_{\theta}$
\begin{subequations}
\begin{eqnarray}
|\uparrow>_\theta & = & \cos(\frac{\theta}{2})|\uparrow> + \sin( \frac{\theta}{2}) |\downarrow>, \\ 
|\downarrow>_\theta & = & -\sin(\frac{\theta}{2})|\uparrow> + \cos( \frac{\theta}{2}) |\downarrow>
\end{eqnarray}
\end{subequations}
Following $t_i$ the time evolution of $\mathcal{W}_i$ again follows Eq. (\ref{hamiltonian}).

With these definitions, the observed value of the quantity in Eq. (\ref{standard}) recorded
by each subsystem will be $(e_1 - e_0)  (e_3 - e_2)$. 

On the completion of all interations
after time $t_N$, each possible history $h$ can be expressed as
a sequence of subhistories of the form
\begin{equation}
\label{histories}
h = ( e_{01}, e_{11},e_{21}, e_{31}, ... e_{0N}, e_{1N}, e_{2N}, e_{3N}).
\end{equation}
The corresponding state vector $| \Psi( h, t_N)>$ has the form
\begin{multline}
\label{state}
| \Psi( h, t_N)> = |\Psi( e_{01}, e_{11}, e_{21}, e_{31}, t_N)>_1 \times \\
 ... | \Psi( e_{0N}, e_{1N}, e_{2N}, e_{3N}, t_N)>_N,
\end{multline}
where each subsystem's states for its 4 possible subhistories are
\begin{subequations}
\label{substates}
\begin{multline}
|\Psi( 0, 1, 0, 1, t_N)> = \frac{1}{\sqrt{2}}\sin( \frac{\theta}{2}) | \uparrow> | \uparrow>_{\theta} \times \\
 | 0, t_N> | 1, t_N> | 0, t_N> | 1, t_N>,
\end{multline}
\begin{multline}
|\Psi( 0, 1, 1, 0, t_N)> =  \frac{1}{\sqrt{2}}\cos( \frac{\theta}{2}) | \uparrow> | \downarrow>_{\theta} \times \\
| 0, t_N> | 1, t_N> | 1, t_N> | 0, t_N>,
\end{multline}
\begin{multline}
|\Psi( 1, 0, 0, 1, t_N)> =  -\frac{1}{\sqrt{2}}\cos( \frac{\theta}{2}) | \downarrow> | \uparrow>_{\theta} \times \\
| 1, t_N> | 0, t_N> | 0, t_N> | 1, t_N> 
\end{multline}
\begin{multline}
|\Psi( 1, 0, 1, 0, t_N)> =  \frac{1}{\sqrt{2}}\sin( \frac{\theta}{2}) | \downarrow> | \downarrow>_{\theta} \times \\
| 1, t_N> | 0, t_N> | 1, t_N> | 0, t_N>.
\end{multline}
\end{subequations}

The weight of any history in the ensemble of initial states, equal by Eq. (\ref{backtofuture}) to its weight in the ensemble of final states, becomes
\begin{multline}
\label{weight}
< \Psi( h, t_N) | \Psi( h, t_N)> = \\ [ \frac{1}{2} \sin^2( \frac{\theta}{2})]^{p(h)} [ \frac{1}{2} \cos^2( \frac{\theta}{2})]^{m(h)}.
\end{multline}
Here $p(h)$ is the count of $i$ in $h$ for which $(e_{1i} - e_{0i})  (e_{3i} - e_{2i})$ is 1, 
and 
$m(h)$ is the count of $i$ in $h$ for which $(e_{1i} - e_{0i})  (e_{3i} - e_{2i})$ is -1.
It follows that each choice of initial state according to Eq. (\ref{weight}) may
be viewed as $N$ independent choices of the random variable 
$(e_1 - e_0)  (e_3 - e_2)$ from an ensemble with probability
$\sin^2( \frac{\theta}{2})$ at value 1 and probability 
$\cos^2( \frac{\theta}{2})$ at value -1. By the central limit
theorem, we then find that for large N, the average over N draws will 
nearly yield
\begin{subequations}
\label{finalresult}
\begin{eqnarray}
< (e_1 - e_0)  (e_3 - e_2) > & = & \sin^2( \frac{\theta}{2}) - \cos^2( \frac{\theta}{2}) \\
& = & -\cos( \theta),
\end{eqnarray}
\end{subequations}
in agreement with Eq. (\ref{standard}).

\section{\label{sec:states}Random State Ensembles}

A paradoxical feature of both states in the time 0 ensemble in Eqs. (\ref{initial}) is that they 
include $| \downarrow>$ components for the spin even though both show 
purely $| \uparrow>$ results for $e_1$ at the first event at time $t_1$.  
What this reflect, we believe, is that the system's initial state contains 
hidden variables the values of which are not directly accessible.
They are determinable only to the extent that they participate in 
events in which they become entangled
with the degrees of freedom of the environment. 
The time 0 ensemble of Eq. (\ref{initial}) also includes linear combinations
of distinct values of the environment variables, but which then assume
single values at the time the entanglement events occurr.
What this implies, we believe, is that the environment variables
do not become part of the macroscopic environmental record until the event
at which they assume their permanently remembered values.
Despite the pecularities of the ensemble in
Eq. (\ref{initial}), however, all that is required of the theory empirically 
is that it show macroscopic results at each event consistent with the Born rule. 
Which we have shown in Section \ref{sec:born} the proposed theory does.

A further limitation restricting access to the initial state's hidden variables
is that the theory proposes to be a complete deterministic description of the universe 
and allows no outside observers.  
The macroscopic experiments which can occurr are only whatever happens to have already
been programmed into the initial state's plan for history.

While the initial state enembles constructed lead
to time evolution without macroscopically ambiguous variables,
the construction is by
a roundabout path. A missing element is 
some criterion which can be applied directly to
candidate initial state ensembles to determine which
of these avoid macroscopically ambiguous time evolutions.
An alternative view of the time evolution of the state vector, however,
partially resolves this problem. 
Rather than taking the state vector as evolving forward in time from time 0,
we could equally well propose it evolves backward from some distant future time $t$.
The initial state for this reverse time evolution is then a random 
draw from an ensemble of $|\Psi( h, t)>$ with corresponding
probability weight $< \Psi( h, t) | \Psi( h, t)>$. Each of these
vectors by construction in Eq. (\ref{entangled1}) will give rise to 
a macroscopically unambiguous time trajectory. 
The problem of finding an initial ensemble then 
becomes finding the decomposition in Eq. (\ref{entangled1}). 
Which task might be facilitated by the
possibility that in the limit of large $t$, as 
the state $| \Psi( h, t)>$ becomes
fully entangled with the rest of the universe, the orthogonality of
the environmental memory components of 
$|\Psi( h, t)>$ and $|\Psi(h', t>$ might become exact for distinct $h$ and $h'$.


\begin{thebibliography}{9}

\bibitem{Everett} H. Everett, Rev. Mod. Phys. 29, 454 (1957).
\bibitem{DeWitt} B. DeWitt, Physics Today 23, 30 (1970).
\bibitem{Zeh} H. D. Zeh, Found. Phys. 1, 69 (1970).
\bibitem{Zurek1} W. H. Zurek, Phys. Rev. D24, 1516 (1981).
\bibitem{Zurek2} W. H. Zurek, Phys. Rev. D26, 1862 (1982).
\bibitem{Schlosshauer} M. Schlosshauer, Rev. Mod. Phys. 76, 1267 (2004).
\bibitem{Wallace} D. Wallace, Studies in the History and Philosophy of Modern Physics 34, 87 (2003).
\end{thebibliography}
\end{document}